\documentclass[aps,prb,twocolumn, showpacs,superscriptaddress,preprintnumbers,amsmath,amssymb]{revtex4-2}

\usepackage[usenames,dvipsnames]{xcolor}

\usepackage[normalem]{ulem}
 \newcommand{\stkout}[1]{\ifmmode\text{\sout{\ensuremath{#1}}}\else\sout{#1}\fi}

\usepackage{float}
\usepackage{graphicx}
\usepackage{dcolumn}
\usepackage{bm}
\usepackage{multirow}
\usepackage{amsmath,amssymb}

\begin{document}
\title{Anisotropic Resonant Scattering from uranium systems at the U $M_4$ edge}

\author{E. Lawrence Bright}
    \affiliation{European Synchrotron Radiation Facility, 71 Avenue des Martyrs, Grenoble 38043, France}
 
 \author{E. N. Ovchinnikova}
    \affiliation{M. V. Lomonosov Moscow State University, Leninskie Gory, Moscow 119991, Russia}
 
 \author{L. M. Harding}
    \affiliation{H. H. Wills Physics Laboratory, University of Bristol, Bristol, BS8 1TL, UK}
 
 \author{D. G. Porter}
    \affiliation{Beamline I16, Diamond Light Source. Harwell Science and Innovation Campus, Didcot, Oxfordshire, OX11 0DE, UK}
    
     \author{R. Springell}
    \affiliation{H. H. Wills Physics Laboratory, University of Bristol, Bristol, BS8 1TL, UK}
 
  \author{V. E. Dmitrienko}
    \affiliation{A. V. Shubnikov Institute of Crystallography, FSRC Crystallography and Photonics RAS, Moscow 119333, Russia}
 
 \author{R. Caciuffo}
    \affiliation{Istituto Nazionale di Fisica Nucleare, Via Dodecaneso 33, IT-16146 Genova, Italy}
 
 \author{G. H. Lander}
   \affiliation{H. H. Wills Physics Laboratory, University of Bristol, Bristol, BS8 1TL, UK}

\date{\today}

\begin{abstract}
We have conducted a series of scattering experiments at the uranium $M_4$ absorption edge on low-symmetry uranium compounds (U$_2$N$_3$ and  U$_3$O$_8$) produced as epitaxial films. At weak and forbidden reflections, we find a resonant signal, independent of temperature, with an energy dependence resembling the imaginary part $f''$ of the scattering factor. Theory, using the FDMNES code, shows that these results can be reliably reproduced assuming that they originate from aspherical 5$f$ electron charge distributions around the U nucleus. Such effects arise from the intrinsic anisotropy of the 5$f$ shell and from the mixing of the 5$f$ electrons of uranium with the outer 2$p$ electrons of the anions. The good agreement between theory and experiment includes azimuthal scattering dependencies, as well as polarization states of the scattered photons. The methodology reported here opens the way for a deeper understanding of the role the 5$f$ electrons in the bonding in actinide compounds.

\end{abstract}


\maketitle

\section{Introduction} \label{Intro}
Diffraction experiments as a function of the incident X-ray energy passing through elemental absorption edges were first performed in the early years of X-ray diffraction \cite{james65}, but became possible on a more expansive scale with the development of synchrotron sources, and were pioneered experimentally by Templeton and Templeton in the 1980s \cite{templeton82,templeton86} and treated theoretically by Dmitrienko in the same period \cite{dmitrienko83,dmitrienko84}. Since that time many experiments have been conducted on different materials, but the vast majority have been at the $K$ edges of transition-metal 3$d$ series of elements. The $K$ edges for these materials span the range from $\sim$ 5 to 10 keV, which are prime energies for both synchrotron sources and diffraction experiments. Much new information about the materials under investigation can be obtained with suitable theoretical understanding \cite{kokubun98,collins01}. However, the $K$-edge has two possible transitions, first the dipole (E1) transition 1$s$ $\rightarrow$ 4$p$, and, second, the quadrupole (E2) transition 1$s$ $\rightarrow$ 3$d$. In the case of 3$d$ metals, these transitions can be almost of equal strength, so difficult to distinguish, although they do occur at slightly different energies. An example of the power of the technique can be seen in the work on TiO$_2$, in which the transitions could be separated and the resulting $p-d$ hybridization of the electronic states identified \cite{kokubun10}. The many ways in which resonant scattering can be observed are discussed in a review article by Kokubun and Dmitrienko \cite{kokubun12}, which also covers work on Ge at the $K$-edge of 11.1 keV.

Other suitable edges for such experiments are the $M_{4,5}$ of the actinides \cite{caciuffo23}. These edges have an energy of 3.55 keV (U $M_5$) to $\sim$ 4.5 keV for Cf, and the E1 transitions represent an electron promoted from the occupied 3$d$ shell to the partially filled 5$f$ shell. In particular, we shall focus on the U $M_4$ edge at 3.726 keV. Diffraction has limitations at these edges, as the wavelength of the incident X-rays is $\lambda$ = 3.327 \AA, which drastically reduces the available reciprocal space that can be examined. The E1 transition (for $M_4$) is 3$d_{3/2}$ $\rightarrow$ 5$f_{5/2}$. This transition is much stronger (as discussed below) than any E2 transitions that involve 3$d$ $\rightarrow$ 6$d$, 6$g$ or 7$s$ states, so it is assumed that the effects measured involve the 5$f$ electrons, which are those of major interest in the actinides. 

The best known E1 transition in this series allows one to probe the magnetic dipole ordering that occurs in many actinide materials. The first experiments to observe this effect were on a single crystal of UAs in 1989 \cite{isaacs89,mawhan90}, and the authors comment that the resonant scattering was about six orders of magnitude greater than any non-resonant scattering in the antiferromagnetic state. Many experiments \cite{caciuffo23} on various aspects of magnetic structures have been explored with this resonant scattering at the $M_{4,5}$ edges of actinides up to and including Pu materials.

In the general case, the anisotropic resonant scattering (ARS) needs to be formulated as a tensor (hence it is often called anisotropic tensor scattering, ATS), and the theory is reviewed in Ref. \cite{caciuffo23} starting with Eq. 55 and continuing to Eq. 62. In the special case of cylindrical symmetry  [i.~e. SO(2)] the main interactions and observables may be represented in a simpler form where the cross sections are given in terms of the two components of polarization of the scattering \cite{hill96}, parallel ($\pi$) and perpendicular ($\sigma$) to the diffraction plane. The results are that the E1 X-ray scattering amplitude contains 3 terms, the first is a non-resonant scalar probing electric charge monopoles. The second term is a rank-1 tensor sensitive to the magnetic dipole moment that, for uranium $M$ edges, gives the large enhancement noted above \cite{isaacs89,mawhan90}. The third term is a rank-2 tensor even under time reversal and sensitive to electric-quadrupole moments and to any asymmetry intrinsic to the crystal lattice.

Many experiments \cite{caciuffo23} have measured the magnetic dipole scattering, which we can conveniently call
E1-$\mathcal{F}^{[1]}$. Such a magnetic term has the characteristic energy dependence of the imaginary part $f''$ of the X-ray form factor and is proportional to the component of the dipole magnetic moment perpendicular to the plane defined by the incident and scattered polarization vectors. The third term in the E1 scattering amplitude, which we call E1-$\mathcal{F}^{[2]}$, has been observed in UPd$_3$ \cite{mcmorrow01}, NpO$_2$ \cite{paixao02,caciuffo03a}, UO$_2$ \cite{wilkins06}, and in their solid solutions \cite{wilkins04}. These results refer to the observation of charge quadrupoles, which cannot be measured by neutron diffraction, and have been of considerable interest \cite{santini09}. They probably exist in more $f$-electron materials than presently realized \cite{suzuki18,lovesey13}. The energy dependence of the scattering in the E1-$\mathcal{F}^{[1]}$ and E1-$\mathcal{F}^{[2]}$ processes is different \cite{santini09} and can be calculated beyond the fast collision approximation \cite{nagao05,nagao06,lovesey12}. For instance, the intensity at the $M_4$ edge in the $\sigma$-$\sigma$ channel for the UO$_2$ (1 1 2) and NpO$_2$ (0 0 3) reflections, due to the E1-$\mathcal{F}^{[2]}$ term, is centered about 2 eV below the position of the magnetic dipole resonance and has an approximate Lorentzian squared shape, contrary to the E1-$\mathcal{F}^{[1]}$ signal that usually exhibits a Lorentzian line shape. However, it must be noted that when the multiplet splitting of the intermediate state can be neglected, an average energy value can be used in the denominator of the E1 scattering amplitude, and the resonant factor can be replaced by a Lorentzian-shaped energy profile.

Similarly, the polarization dependencies are different for E1-$\mathcal{F}^{[1]}$ and E1-$\mathcal{F}^{[2]}$. In the former the incident $\sigma$ polarization is all rotated to $\pi$ radiation, whereas in the latter process both $\sigma$-$\sigma$, and $\sigma$-$\pi$ polarizations exist. The azimuth angle dependence of the resonant Bragg peaks (the variation of the peak intensity while the sample is rotated about the scattering vector) of both cross sections provide information on the mutual orientations of the aspherical electronic clouds in the crystallographic unit cell. As well as electric quadrupoles, the E1-$\mathcal{F}^{[2]}$ scattering also occurs when the magnetic structure has at least two components that are non-collinear, i.~e. either 2\textbf{k}, or 3\textbf{k} magnetic configurations \cite{longfield02}.

\begin{figure}
\centering
\includegraphics[width=1.0\columnwidth]{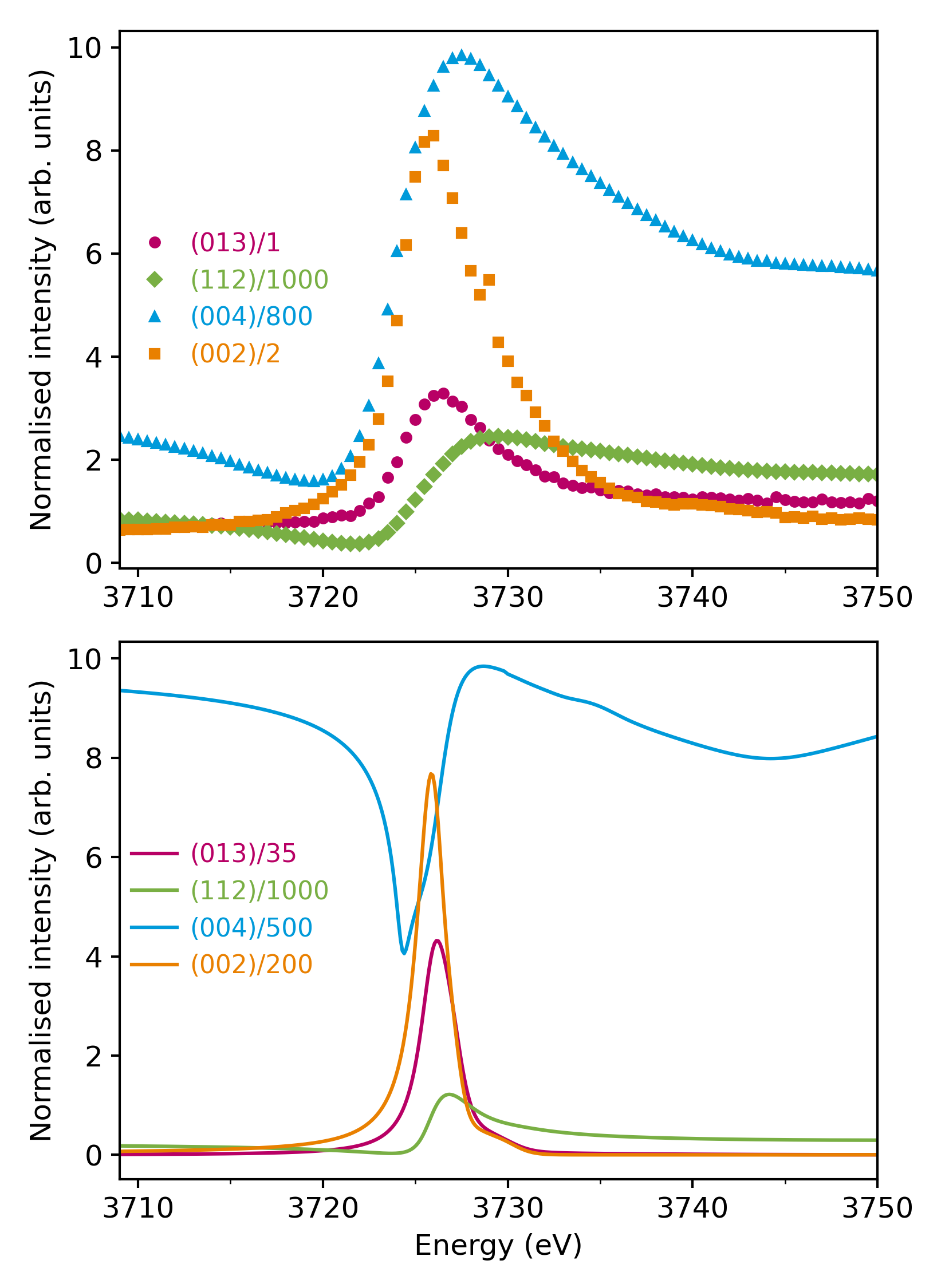}
\caption{The upper panel shows experimental results (no absorption correction or background subtraction) for the strong (004) and (112), for the weak (002), and the forbidden (013) reflections in U$_2$N$_3$. The profiles are independent of temperature.
The lower panel shows the theoretical energy profiles of the same four reflections without any broadening due to the experimental resolution.}
\label{endepU2N3}
\end{figure}

We have discussed the more conventional resonant scattering as performed in the actinides in some detail in order to make a contrast with the results reported in the present paper. Our first observation was reported briefly in 2019 using epitaxial films of the cubic $bcc$ U$_2$N$_3$ \cite{lawrence19}. We shall discuss reflections from this material in more detail later, but we show in Fig. \ref{endepU2N3} the energy dependence for various reflections measured with this material.

The shape of the energy curves closely follows that of the E1 resonance anticipated from the $f''$ term in the cross section. The position and shape of the peak in energy strongly suggest this is an E1 process. The energy dependence obtained from theory (see Sec. \ref{theory}) is compared with experimental results in Fig. \ref{endepU2N3}. The calculated curves are clearly narrower than the experimental ones, but this is a question of experimental resolution. The overall agreement is excellent.

U$_2$N$_3$ also orders antiferromagnetically (AF) at T$_N$ $\sim$ 75 K. Evidence for this is reported in Ref. \cite{lawrence19}. The new AF reflections appear at non-$bcc$ reciprocal lattice points, i.~e. at reflections with $h$ + $k$ + $\ell$ = odd, which indicates that in the AF state the dipole moments related by the $bcc$ operator have oppositely directed moments. The exact AF configuration is unknown, but it is important to stress that the effects reported in the present paper are unrelated to the AF order. First, the effects have been observed on purely charge-related reflections, i.~e. $h$ + $k$ + $\ell$ = even, and, second, no temperature dependence is found for any of the effects discussed here.

There is \emph{ample} evidence, especially in the study on U$_2$N$_3$, that the effects are due to anisotropic 5$f$ electron charge distributions. These will become evident at absent or weak Bragg reflections when the spherical charge distribution due to the radon core (86 electrons) is subtracted due to the out-of-phase contributions from two different uranium atoms. In the studies reported below we have such conditions in the unit cell. The effects we observe can then be seen when the spherical core distributions are subtracted, and the remaining part represents the difference between the anisotropic charge distributions from the 5$f$ states. The fact that these have a maximum value at the $M_4$ absorption edge, is simply a consequence of the maximum of the $f''$ component at this energy. They unambiguously assign the effects to aspherical 5$f$ distributions, possibly associated with covalency, presumably (in the case of U$_2$N$_3$) between the U 5$f$ states and the nitrogen 2$p$ states.

\begin{figure}
\centering
\includegraphics[width=1.0\columnwidth]{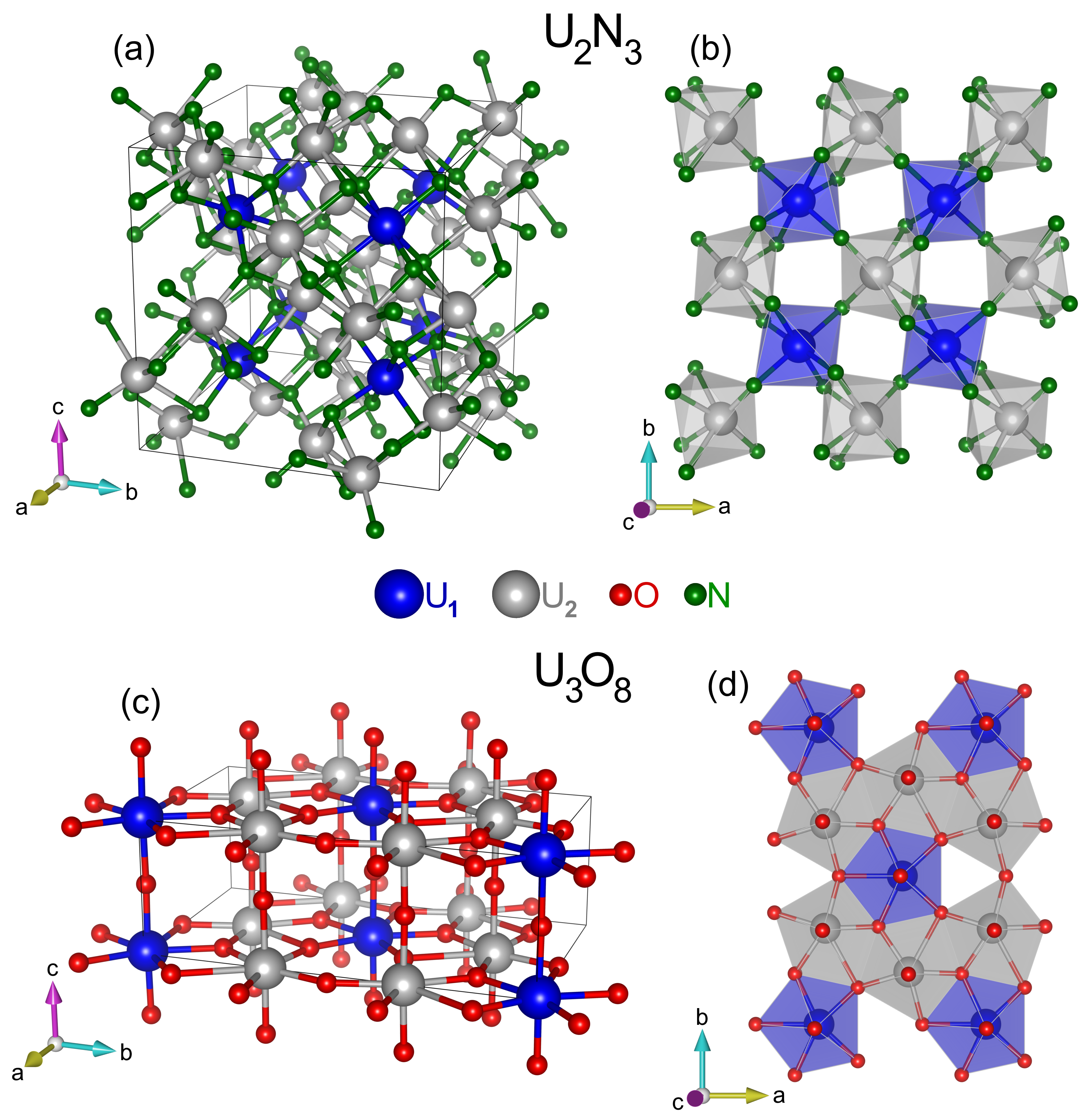}

\caption{Crystallographic structures of U$_2$N$_3$ (panels a and b) and U$_3$O$_8$ (panels c and d). The two independent uranium sites, U$_{1}$ and U$_{2}$, in each material are indicated as dark blue and silver spheres.}
\label{crystalstructure}
\end{figure}

We show in Fig. \ref{crystalstructure} the structures of the two uranium compounds we have examined. In both cases there are two independent sites for the U atoms, and it is the differences in the charge distributions between these two sites that gives rise to the anisotropic resonant scattering.

These effects cannot be observed in reflections that are absent due to global symmetry constraints (e. g. at positions forbidden by $fcc$ or $bcc$ symmetry operators), but can be present at forbidden reflections \cite{collins01,kokubun10} due to glide-plane operators. They can also be present at weak reflections, where contributions from the uranium atoms in the unit cell are out of phase. The effect cannot be observed in high-symmetry structures such as UO$_2$ ($fcc$ CaF$_2$ structure) or UN ($fcc$ NaCl structure). Even in the well-known compound URu$_2$Si$_2$ with the I4/mmm (SG no. 139) tetragonal structure, the effect will not be present, as there is only one U atom at the origin of the unit cell.

\section{Experimental details}
Measurements were performed on epitaxial thin films fabricated at the FaRMS facility at the University of Bristol, UK \cite{springell22} which has a dedicated actinide DC magnetron sputtering system.

Epitaxial thin films provide a series of advantages for the measurements performed in this study. Firstly, they allow easy fabrication and stabilization of single crystals such as U$_2$N$_3$, which has not previously been produced in bulk. Secondly, the low volume of radioactive material allows for easy handling and transportation of the samples. Thirdly, to facilitate a major aim of the experiments of obtaining azimuthal scans, where the sample is rotated about the scattering vector, and the intensity determined as a function of $\Psi$, the so-called azimuthal angle. The major experimental difficulty is associated with a large absorption of X-ray beams of this tender energy incident on a sample containing uranium. As given in Ref. \cite{caciuffo23} the attenuation length (1/e) of such beams at the $M_4$ edge into uranium metal is $\sim$ 400 nm, somewhat longer for an oxide with lower density. A large, flat surface (5 $\times$ 5 mm$^2$) with a thickness of $\sim$ 200 nm gives a uniform scattering volume as it is rotated about the scattering vector to perform the azimuthal scan, provided also that the angle to the specular direction of the film is less than $\sim$ 20 deg. Whereas qualitative results are relatively easy to obtain, quantitative results for the azimuthal intensity that can be compared to theory are much more difficult to extract. 

Films of U$_2$N$_3$ and U$_3$O$_8$ were deposited by sputtering in N$_2$ and O$_2$ partial pressures, respectively, as described previously \cite{lawrence18,lawrencebright22}. To avoid oxidation, all films were covered by a polycrstalline cap ($\sim$ 50 nm) of Nb. 

U$_2$N$_3$ was deposited on (001) oriented CaF$_2$, producing U$_2$N$_3$ with the principle axes aligned in the specular direction. Due to the symmetry of the U$_2$N$_3$ bixbyite structure, with non-equivalent $a$, $b$, and $c$ axes and a [111] screw axis, this effectively produces two domains. These domains will have completely overlapping Bragg reflections. For convenience, we will describe the film with the [001] axis specular, with the two domains defined by having either the [100] or [010] axis along the CaF$_2$ [100] direction. 

U$_3$O$_8$ was also deposited on a (001) CaF$_2$ substrate, producing a film with 8 domains with [131] specular. Non-specular reflections of domains do not overlap, making them easy to distinguish.

ARS experiments were performed using the I16 diffractometer \cite{I21} at the Diamond Synchrotron (UK). The energy of the incident X-ray beam has been tuned to the uranium $M_4$ edge at 3.726 keV.
All the results in this paper refer to the samples at \textit{room temperature}. In Ref. \cite{lawrence19} tests were done on a forbidden reflection of U$_2$N$_3$ as a function of temperature, and no $T$-dependence was found. We have assumed that these effects are associated with bonding in the material, and thus no $T$-dependence is expected. 

It is also important to determine whether the polarization of the scattered radiation is unrotated, i.~e. $\sigma$-$\sigma$ or rotated, i.~e. $\sigma$-$\pi$, which is measured in standard fashion by using an Au (111) crystal as an analyzer before the detector. Since the results reported here are of weak intensities, and the use of an analyzer reduces the observed signal, we have only performed limited polarization scans. 

A major further difficulty is that there are domains in all of the films. These have been studied and characterized at Bristol before the synchrotron experiments. Multiple scattering is also a possibility.

\section{Theory} \label{theory}
Our studies of resonant X-ray scattering in U$_2$N$_3$ and U$_3$O$_8$ at the incident radiation energy close to the $M_4$ absorption edge demonstrated strong anisotropy of resonant atomic factors of uranium corresponding to the E1 transitions between the 3$d_{3/2}$ and virtual 5$f$ states. The study of the spectral shape of both the forbidden reflection (105), and several weak allowed reflections in U$_2$N$_3$ and U$_3$O$_8$, has shown that their spectral shape has a form of a peak close to the $M_4$ absorption edge, implying that the resonant contribution to the atomic factor is sufficiently strong in comparison to the charge scattering, in contrast to the situation at the $K$-edge \cite{mukhamedzhanov75}.  The pronounced azimuthal dependence confirms this, as well as the existence of a scattering channel with a change in polarization. In both studied crystals, uranium atoms occupy two crystal sites with different local symmetry, hence the spectral and angular properties of reflections are determined by the interference of the waves scattered by non-equivalent atoms and by the electronic density. This makes the azimuthal dependence of reflections dependent on energy. Such a phenomenon was also observed in Fe $K$-edge
\cite{beutier09}. 

\begin{figure}
\centering
\includegraphics[width=1.0\columnwidth]{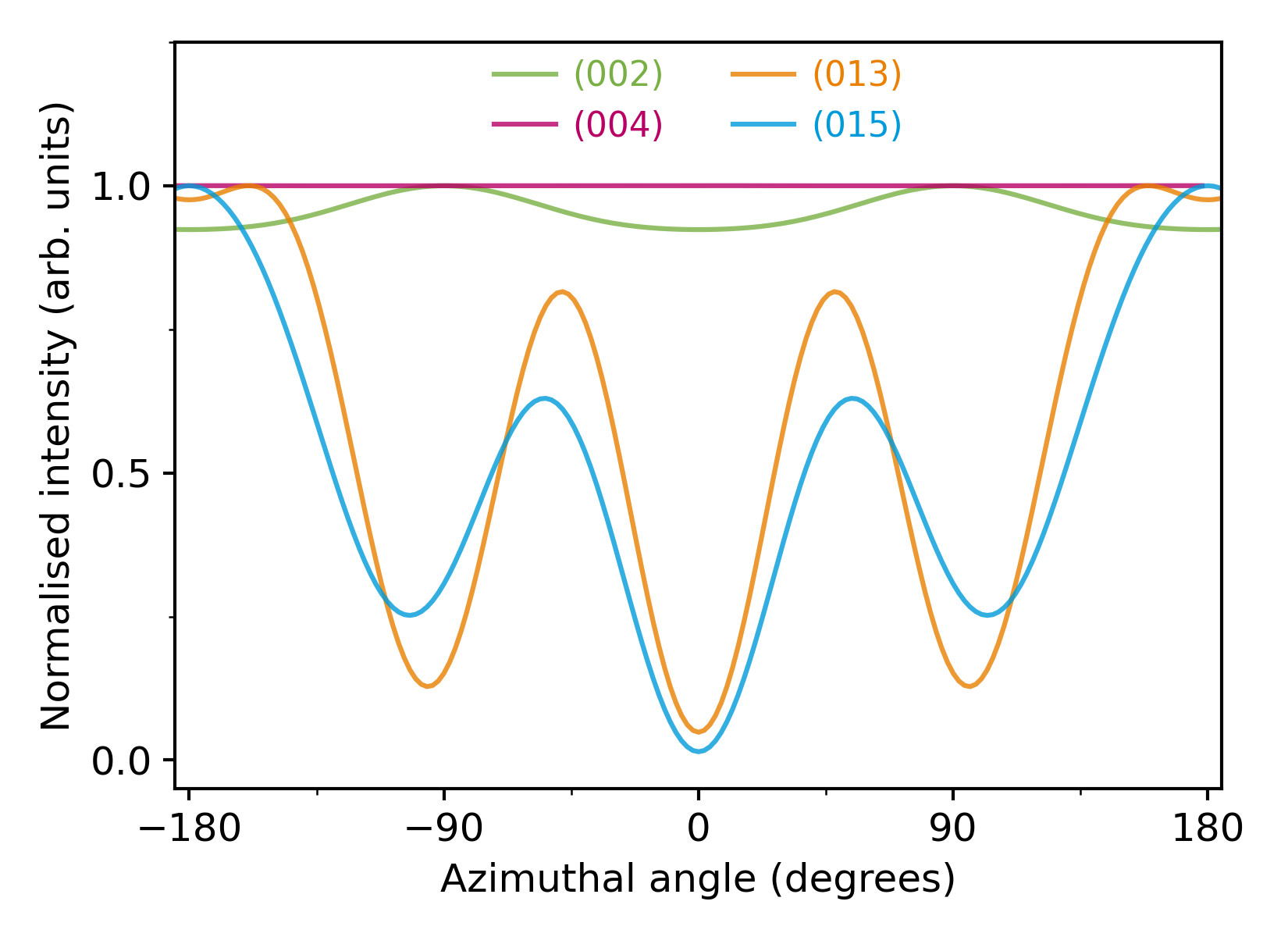}
\caption{Calculated azimuthal dependence of various reflections in U$_2$N$_3$. Note that the (004) is the strongest pure charge reflection, the (002) is a weak reflection that has an additional ARS contribution, and the (013) and (015) reflections are forbidden. The latter two have only ARS scattering. The maximum of each intensity has been normalized to unity.  
}
\label{Fig2t}
\end{figure}

	In U$_2$N$_3$ the local symmetry of the U$_{1}$ atom is \={3}, hence the non-magnetic dipole resonant atomic factor is uniaxial with two independent components, whereas the atomic factor of the U$_{2}$ atom with 2 local symmetry is not uniaxial and possesses 3 independent components. In U$_3$O$_8$ the atomic tensor factors of both U$_{1}$ and U$_{2}$ are not uniaxial, but their symmetry differs from one other.  All tensor components have their specific spectral shapes, providing a variety of spectral and azimuthal properties of resonant reflections, which are determined by their combinations. 
We will not describe in detail all the features of the tensor factors of uranium, but we will demonstrate some statements using the example of calculations performed with the FDMNES program \cite{bunau09}.  It allows us to make a variety of calculations, including calculating energy spectra and azimuthal dependences of reflections, and makes it possible to vary many physical parameters that describe the system under study for comparison with experimental data.

Fig. \ref{Fig2t} shows calculations of the azimuthal dependence of various reflections in U$_2$N$_3$. There is, of course, also a dependence on the intensity of the energy displacement from the edge but the azimuthal symmetry is largely independent of this factor.
The large variety of shapes of the azimuthal dependences is due to the difference in the spectral shape of the components of the tensor atomic factor for each uranium atom, which contributes to individual reflections, as well as the type of interference of waves scattered by atoms of positions U$_{1}$ and U$_{2}$. The (004) reflection, which is the strongest in the structure has, of course, no azimuthal dependence and is all $\sigma$-$\sigma$.

\begin{figure}
\centering
\includegraphics[width=1.0\columnwidth]{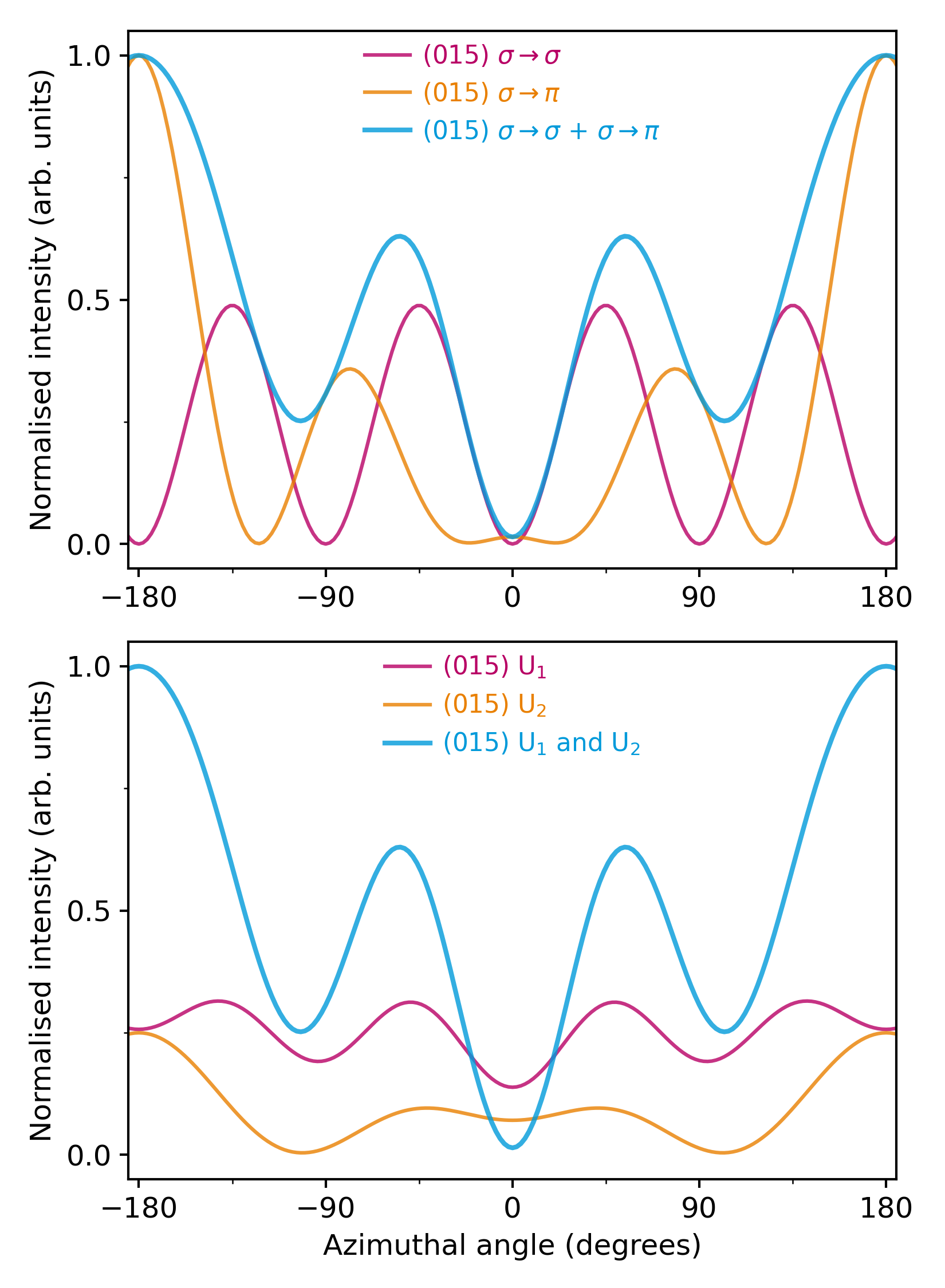}
\caption{Azimuthal angle dependence of the square of the modulus of the structural amplitude for the (015) reflection. Upper panel: $\sigma$-$\sigma$ channel (magenta line),
$\sigma$-$\pi$ (orange line) channel, and the sum of the two polarization channels (cyan line). Lower panel: calculated curves taking into account the contribution only from atoms of position U$_{1}$ (magenta line) and only from atoms of position U$_{2}$ (orange line), as well as taking into account both positions of uranium (cyan line).}

\label{Fig3t}
\end{figure}

Fig. \ref{Fig3t} gives further details of the (015) reflection. The upper panel shows the azimuthal dependence of the the square of the modulus of the structural amplitude for the
$\sigma$-$\sigma$ and
$\sigma$-$\pi$ channels, together with their sum. It is worth noting here that the $\sigma$-$\sigma$ intensity is zero at the azimuth where the total intensity has a maximum, so there should be a strong $\sigma$-$\pi$ contribution at this point, which was found experimentally. The lower panel of Fig. \ref{Fig3t} shows the same quantity taking into account the contribution only from atoms of position U$_{1}$ (magenta line) and only atoms of position U$_{2}$ (orange line), as well as when taking into account both positions of uranium (cyan line). Note that the cyan curve is not the sum of the other two, since it is the square of the modulus of the sum of the scattering amplitudes from the two uranium positions, taking into account the phase difference.

The situation is even more complicated for the non-forbidden reflection, because it is necessary to take into account the charge scattering, which participates in the interference of the waves. There is a good chance to separate the resonant and charge scattering using polarization analysis, because the latter forbids  the $\sigma$-$\pi$ scattering channel \cite{ovchinnikova19}.  Calculations demonstrate strong difference of the azimuthal dependences of the  $\sigma$-$\pi$ and $\sigma$-$\sigma$ scattering. In particular, strong  $\sigma$-$\pi$ scattering is expected for forbidden (103) and (105) reflections, and this has been confirmed experimentally (Fig. \ref{az105}), but for allowed reflections $\sigma$-$\sigma$ is stronger than $\sigma$-$\pi$.

\section{Results and Discussion}
\subsection{U$_2$N$_3$}
This material has the body-centered cubic bixbyite structure common to materials such as Mn$_2$O$_3$, which has an inversion center at (000). Space group  no. 206 Ia\={3}. Because the film (200 nm) is deposited on a CaF$_2$ substrate, there is some small strain (1.9\%), the c axis = 10.80 \AA~in the growth direction, and the basal plane axes are 10.60 \AA. We have performed DFT simulations to see whether this small strain, which results in an orthorhombic structure, changes significantly the symmetry conditions of the uranium atoms, but they show that the effects are very small. We therefore keep the cubic $bcc$ structure as a good approximation to the symmetry in the film. Orientation [001] vertical, \textit{a} and \textit{b} in plane.

As discussed in Sec. \ref{Intro}, U$_2$N$_3$ also orders magnetically at $\sim$ 75 K, see Ref. \cite{lawrence19}. The AF order gives rise to new reflections at positions $h$ + $k$ + $\ell$ = odd, whereas all measurements reported here have been made at true bcc positions, i.~e. $h$ + $k$ + $\ell$ = even, and are at room temperature.

There are two types of uranium in the unit cell: U$_{1}$ sits at 8b position, point symmetry (.\={3}.) with coordinates ( ¼ ¼ ¼) and this atom is at an inversion center. The second uranium U$_{2}$ sits at position 24d with coordinates (x 0 ¼) with x $\sim$ - 0.02 and there is no inversion center at this site, the point symmetry is (2 . .). 

There is no 4-fold symmetry element in this space group. This implies that the [100] and [010] axes are different. In turn, this implies that there are two domains in the film with an [001] axis as the growth direction. To compare theory and experiment, we need to average over the two domains. In practice, what appears to be a theoretical curve for the (002) with a repeat of 180 deg in the azimuthal angle, will result in two patterns displaced by 90 deg, so the overall repeat appears to be 90 deg in the azimuthal. For other reflections the domain averaging is more complex.

\begin{figure}
\centering
\includegraphics[width=1.0\columnwidth]{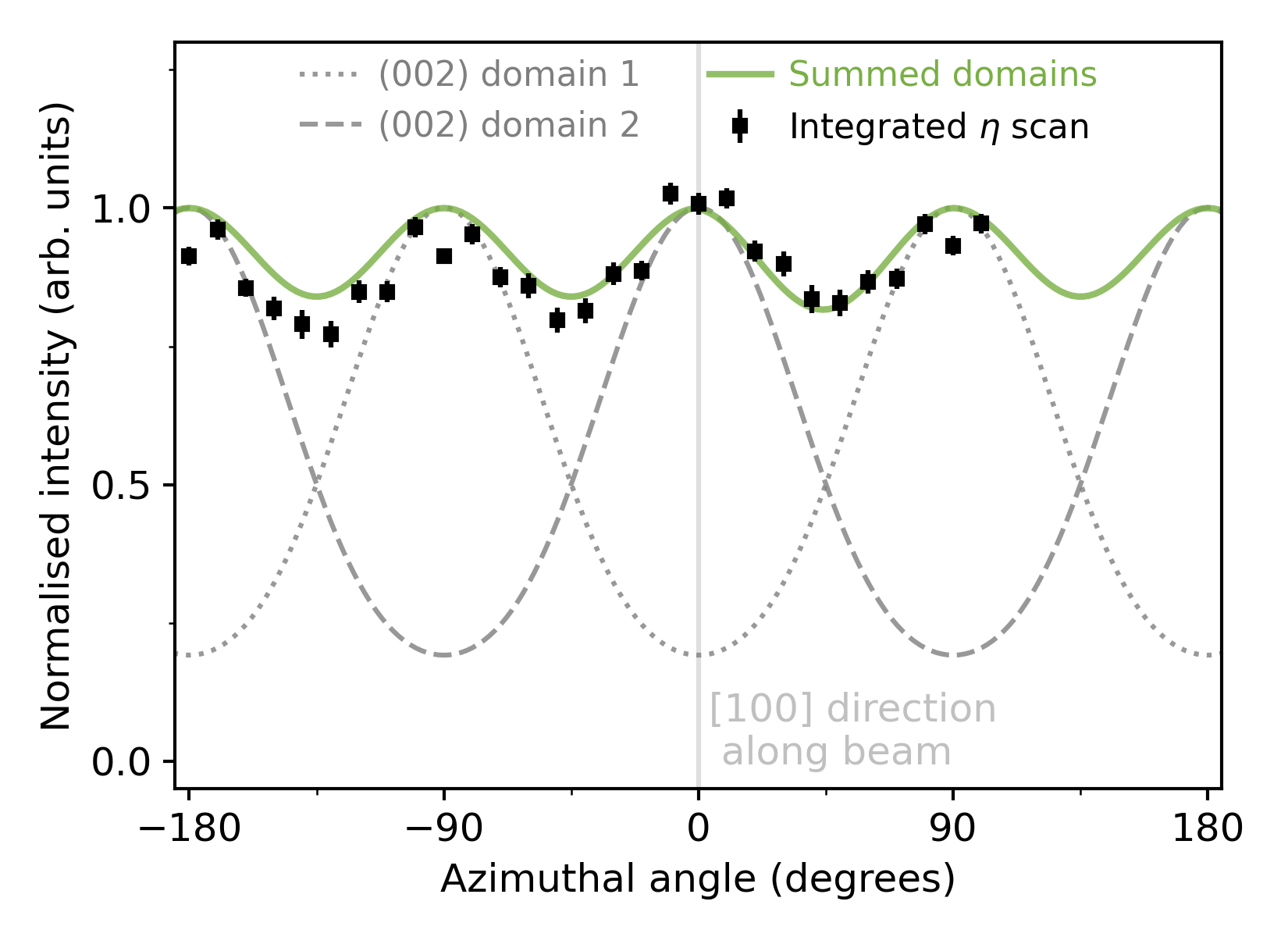}
\caption{Observed azimuthal dependence of (002) reflection from U$_2$N$_3$. The dashed curves show the theory for two domains, which are out of phase by 90 deg. The green curve gives the predicted sum. The (002) reflection is allowed, but has a weak intensity of $\sim$ 0.2 on this scale. The allowed reflection has no azimuthal dependence.}
\label{az002}
\end{figure}

Results for azimuthal scans for the (002) allowed (but weak) reflection are shown in Fig. \ref{az002}. Polarization scans showed the majority scattering was in the $\sigma$-$\sigma$ (unrotated) channel, but since the reflection is also allowed this is not surprising. We did find a small signal in $\sigma$-$\pi$, consistent with theory.

\begin{figure}
\centering
\includegraphics[width=1.0\columnwidth]{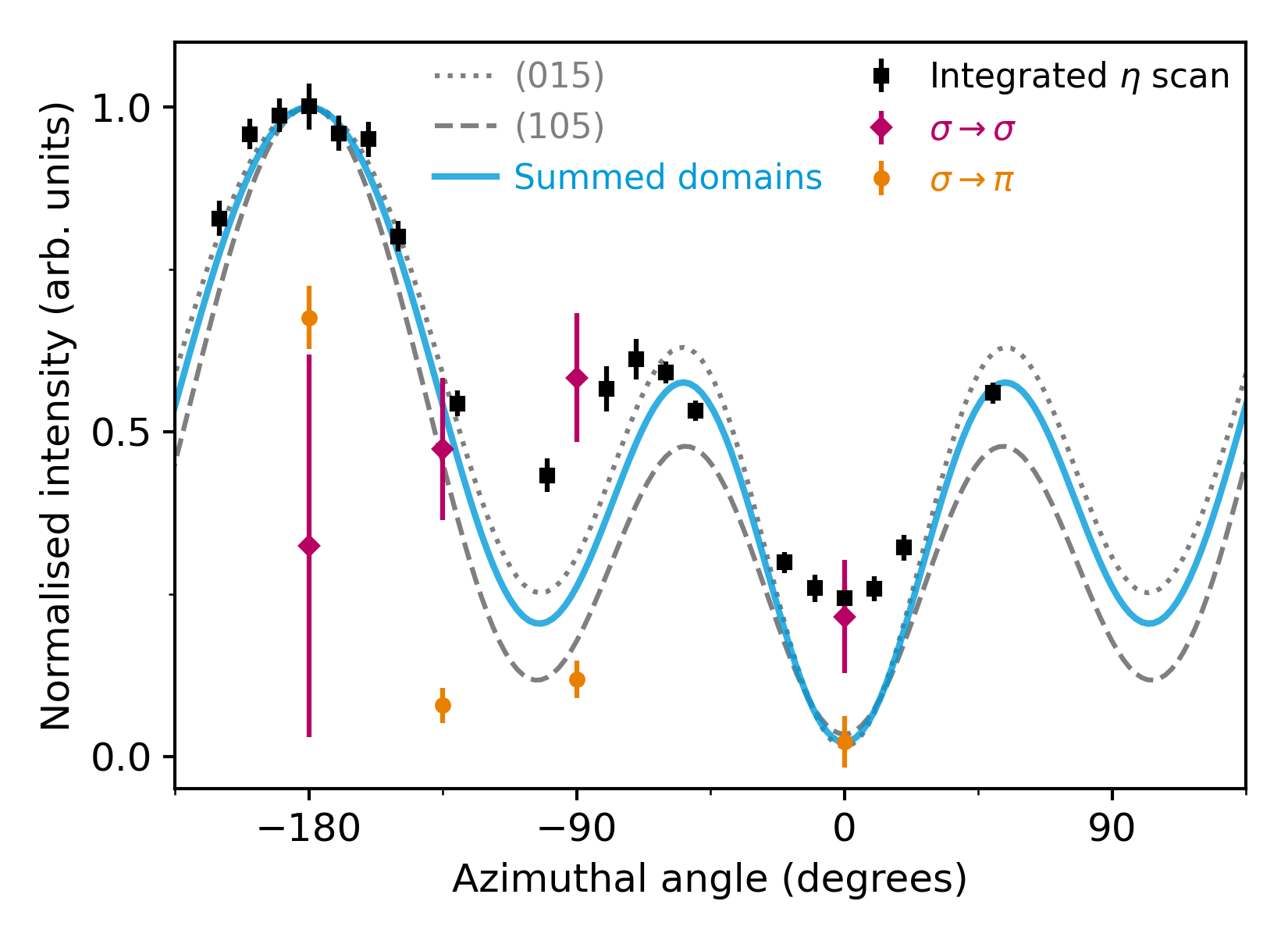}
\caption{Integrated intensities (black squares) as a function of the azimuthal angle $\Psi$ compared to theory calculations where the results are a sum of the theory for (105) and (015). 
Orange circles indicate $\sigma$-$\pi$ contribution and red diamonds indicate $\sigma$-$\sigma$. The theoretical intensity for the (105) and (015) are shown as dotted and dashed lines, respectively, and the half of their sum as the solid red line. Theory and experiment are normalised at the maximum value.
}
\label{az105}
\end{figure}

We now turn to the forbidden reflections (105) and (015). We have already shown in Fig. \ref{endepU2N3} the energy dependence of the intensity found at (013), which like the (015) is forbidden in this space group due to the presence of a glide plane. For azimuthal scattering we have chosen the (105) as the angle to the specular (11.3 deg) is smaller than for the (103). We also have similar theoretical curves for the (105) and (015), see Fig. \ref{Fig3t} (lower panel).
Recall that the domains will result in a summing of these two reflections before we can compare experiment with theory. 
Figure \ref{az105} shows the experimental results compared to theory for the (105) + (015) domains.

Experimentally we find a minimum intensity at $\Psi$ = 0 (when the [100] is along the beam direction), but it is not zero. Despite these scans all being $\eta$ scans (i.~e. the film is rocked through the reciprocal lattice point) some small intensity ($\sim$ 0.3 on scale of Fig. \ref{az105}) remains and this we ascribe to background multiple scattering. Notice here that the only position where we have found appreciable rotated (i.~e. $\sigma$-$\pi$) scattering is at the position of the maximum. This is predicted by theory (see Fig. \ref{Fig3t}, lower panel) and the agreement with experimental results is clearly acceptable. A further test was made by rotating the sample 90 deg and the minimum in scattering rotated by the same angle. 

Approximately, the intensity of the ARS scattering is between two and three orders of magnitude lower than the strong Bragg reflections from the structure, which is also in agreement with theory.

\subsection{U$_3$O$_8$}
U$_3$O$_8$ is an important product of the oxidation of UO$_2$. The structure of the $\alpha$-form is orthorhombic. Although there is a tendency to give the space group as no. 38 with symmetry Amm2, this loses the connection to the hexagonal high-temperature form with Space Group no. 189 and P\={6}2m symmetry. We have therefore found it easier to retain this connection by defining the orthorhombic form with the symmetry C2mm and lattice parameters (at RT) of a = 6.715 \AA, b = 11.96 \AA, and c = 4.15\AA. When this converts to the hexagonal form, the c axis remains the same, and there is simply a shift of the atoms in the \textit{ab} plane. This is consistent with the early work on the crystal structures reported by Loopstra \cite{loopstra70}. More recent work has tended to use the description in terms of the Amm2 notation
\cite{miskowiec20,miskowiec21,sainz23}. At ~ 25 K this material orders antiferromagnetically \cite{miskowiec21,sainz23}, but we have examined the thin film sample only at room temperature.

The symmetry of this system is low, there are two different U positions in the unit cell.  U$_{1}$ is at the position (x 0 0), with x = 0.962 on a 2-fold axis. This atom is supposed to have a U$^{6+}$ valence state, so there should be no 5$f$ electrons associated with U$_{1}$, as the 5$f$ shell is empty. However, transitions from the core 3$d$ states into the empty 5$f$ shell are still possible. This interpretation is consistent with a recent study with resonant inelastic X-ray scattering \cite{lawrencebright22}.

The 4 U$_{2}$ atoms, with valency U$^{5+}$, i.~e. 5$f^1$ are at positions (x y 0) with x = 0, y = 0.324 and they sit on a mirror plane. The space group is non-centrosymmetric, and neither U atom is at positions of inversion symmetry. There are no forbidden reflections in this system (except that $h$ + $k$ = even from the C-face centering). No extra scattering was found on reflections $h$+$k$ = odd.

\begin{figure}
\centering
\includegraphics[width=1.0\columnwidth]{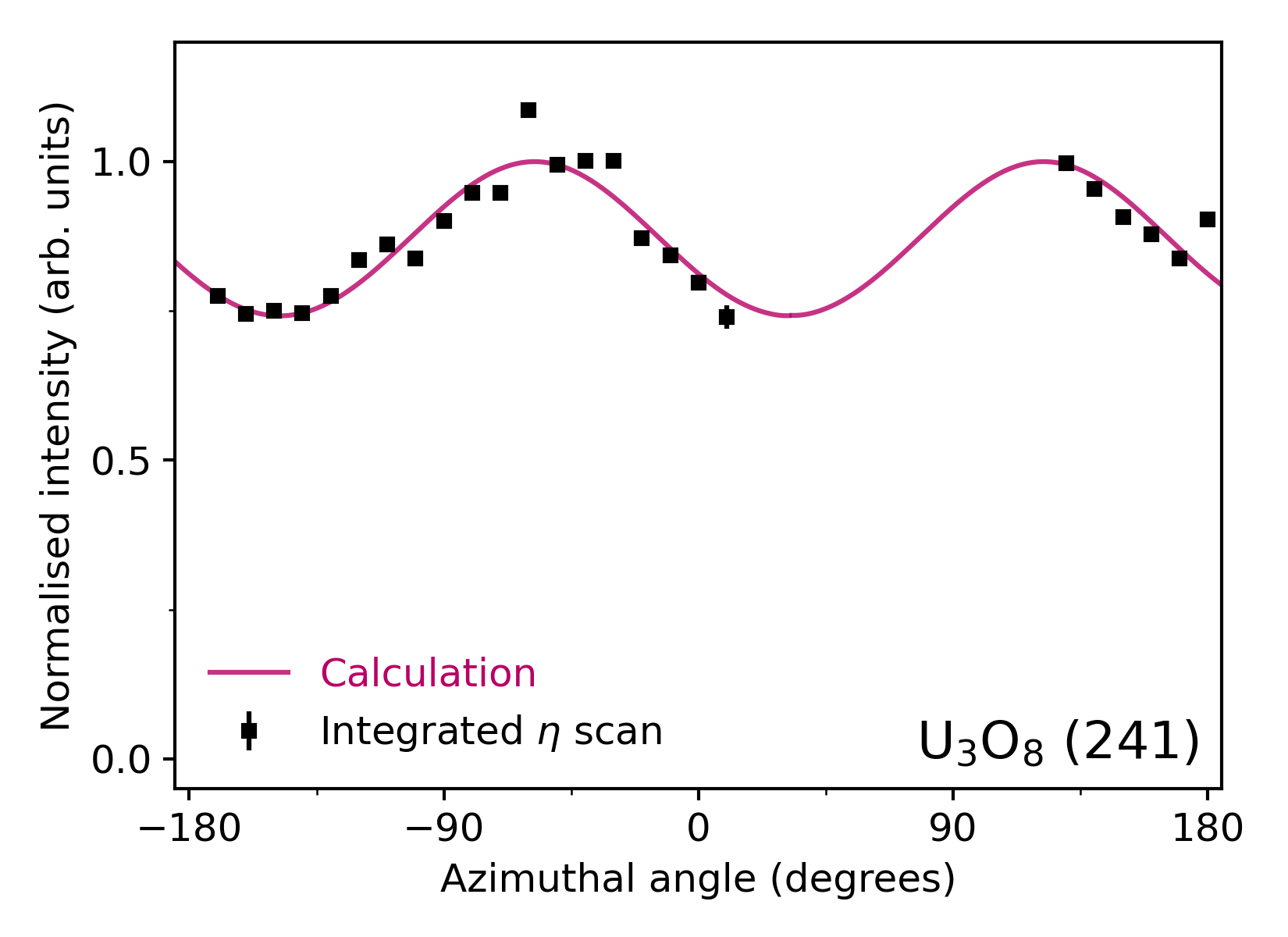}
\caption{Integrated intensities (black squares) as a function of the azimuthal angle $\Psi$ of the (241) reflection from U$_3$O$_8$. Calculated values are shown by the solid line. 
}
\label{az241}
\end{figure}

It proved difficult to measure azimuthal dependencies, because of the need to make large absorption corrections as a function of $\Psi$. We established that the weak (241) reflections, with an angle of 14.5 deg to the specular had the azimuthal dependence shown in Fig. \ref{az241}. With a repeat of 180 deg. The allowed (241) reflection has a calculated intensity 3.0\% of the strongest reflection (001). So, the forbidden intensity is $\sim$ 1\% of strongest reflection.

Extra energy dependent contributions were found for a number of other weak reflections, but their azimuthal dependence was not readily established. Only a very small contribution ($<$ 5\%) was found in the $\sigma$-$\pi$ channel, so the $\sigma$-$\sigma$ dominates. There are a number of domains in this system, but they do not align with the principal domain we have chosen, so there is no overlap in comparing with theory. The latter gives a repeat of 180 deg with only small  $\sigma$-$\pi$ cross section, which is consistent with the experiment.

\section{Conclusions}
The experiments and theory presented here show clearly that the uranium atoms in the investigated structures exhibit aspherical 5$f$ charge distributions. 

The results obtained show that in U$_2$N$_3$ the contribution to resonant scattering from atoms U$_{1}$ is significantly less than from atoms U$_{2}$, but it cannot be neglected. This conclusion was deduced from intensity considerations in Ref. \cite{lawrence19}, but lacked quantitative evidence from azimuthal scans. The present theory confirms that this is the case. The tensor atomic factor of atoms U$_{1}$ has uniaxial symmetry, which is not the case for U$_{2}$.

In the case of U$_3$O$_8$ both uranium sites contribute to the ARS of the reflections. The U$_{1}$ site in this material is U$^{6+}$ so has no occupied 5$f$ states, however, the ARS cross section depends on the status and asphericity of the unoccupied 5$f$ states.

An interesting paper by Lovesey \cite{lovesey23} has suggested that we may be observing uranium octupoles in U$_2$N$_3$, which would require an E2 transition at the $M_4$ edge. Given our discussion in the Appendix about E2 transitions, together with the strong evidence for a dipole (E1) transition in the energy dependence (shown in Fig. \ref{endepU2N3}), we believe this interpretation \cite{lovesey23} is unlikely, and too small to be observed even if present.

Based on a rough estimate of the ARS diffraction intensity from these two systems, the ratio to the strongest reflections from the crystal structures is between 0.1 and 1\%. This is not a particularly difficult limit with synchrotron sources, although measuring accurately the azimuthal dependencies is more challenging due to multiple scattering, as well as the large absorption at the resonant energy. 

Quite possibly, many more such systems can be found and measured to give further evidence for these effects. The bulk of the data should be able to be modelled to examine the orbital occupation of the 5$f$ states around the U nuclei; thus, giving a more quantitative understanding of the covalency in these materials. For example, the crystal truncation rod experiments of Stubbs \textit{et al} on UO$_2$ \cite{stubbs15} could be combined by measuring at the U $M_4$ edge with dissolution experiments (Springell \textit{et al.} \cite{springell15}) to search for complexes involving uranyl-based (U$^{6+}$) deposits on the surface. \textit{Ab initio} calculations, such as those by Arts \textit{et al.} \cite{arts24} could then model the molecular structure to understand better what happens at the atomic level during dissolution.

The presence of aspherical 5$f$ orbitals largely depends on the symmetry of the lattice and covalent interactions. However, our current experiment does not allow us to determine the specific orbitals involved in covalency or the extent of it.  Advanced \textit{ab initio} electronic structure calculations, such as those combining density functional theory and its time-dependent extension, or dynamical mean field theory, are required \cite{kaltsoyannis24}. Nevertheless, it is only by gathering more experimental information,  as done in the current work by elastic resonant X-ray scattering, or by spectroscopy techniques \cite{caciuffo23,kvashnina17,kaltsoyannis24}, that a precise model of the covalency in actinide bonding can be established.

\section{Acknowledgment}
We would like to thank Steve Collins, Alessandro Bombardi, and Gerrit van der Laan for discussions. This work was carried out with the support of Diamond Light Source, instrument I16 (proposals MM27807, MM34651).

These experiments were started in 2018, with a report of the first observation of ARS from U$_2$N$_3$ published in 2019 \cite{lawrence19}. More experiments were then performed in 2021, and the theoretical calculations were completed by the end of 2021. A further experiment to test the theoretical predictions was performed in 2023.

\section{Appendix. Discussion of possible higher-order transitions in RXES  at the uranium $M$ edges.}

The E2 contributions at the $M_{4,5}$ edges of actinides indirectly probe 5$f$-shell higher order multipoles (up to rank 4), through 3$d$ $\rightarrow$ (6$d$, 6$g$, 7$s$) transitions, or intermultiplet processes. Up to now, there has been no definitive identification of the E2 term at the actinide $M$ edges due to the involvement of intermediate states that are delocalized orbitals, resulting in weak overlap integrals and intensities much weaker than the E1 transition. We should emphasize here that the absence of an observable E2 transition at the actinide $M_{4,5}$ edges contrasts with the transitions at the $K$ edges of transition metals, where the E1 and E2 transitions are of similar magnitude.
This is because the E2 transition at the $K$-edge connects two relatively localized orbitals (1$s$ $\rightarrow$ 3$d$) and is comparable in intensity, or even stronger, than the E1 transition, which connects the 1$s$ state with the more delocalized 4$s$ and 4$p$ states.

A quantitative estimate can be made by referencing a paper on URu$_2$Si$_2$\cite{wang17} and examining the terms included in the cross sections of such transitions. Using the formulae given as Equations (32) and (33) in Ref. \cite{wang17}, we can estimate the relative strength of the ($M_4$, E1) transition compared to that of the ($M_4$, E2) transition, i.~e.

\begin{equation}
\frac{I(M_4, E2)}{I(M_4, E1)} \propto \Big( \frac{\langle 3d|r^2|6d, 6g, 7s \rangle}{\langle 3d|r|5f \rangle}\Big)^4
\label{E2overE1}
\end{equation}

\noindent

where the wave-vector (k) and core-hole lifetimes ($\Gamma$) given in Ref. \cite{wang17}, Eq. (32, 33), drop out, as we are examining the same $M$ edges in both cases. The 4$th$ power emerges because the process involves a photon in/photon out, and the amplitude encompasses the ground and intermediate state, followed by the reverse process. This results in the square of the matrix element. For intensity, this requires the 4$th$ power. We use the values
$\langle 3d | r | 5f \rangle$ = - 0.04452, $\langle 3d | r^2 | 6d \rangle$ = 0.00147, and
$\langle 3d | r^2 | 7s \rangle$ = – 0.00047 \cite{vanderlaan}, so the ratio for the 6$d$ transition is down a factor of 30, and that for the 7$s$ is down a factor of 95. It is the 4$th$ power of these numbers that results in factors of less than 10$^{-5}$, so E2 transitions at the $M_{4,5}$ edges will be difficult to observe, consistent with the fact that none have been observed so far. It is the relatively large value of the overlap between the wavefunctions 3$d$ and 5$f$ in the actinides that gives rise to the large E1 term, and the enhancements reported in such measurements \cite{isaacs89,mawhan90}.

\bibliography{ARS}

\end{document}